# Magnon-drag thermopower and Nernst coefficient in Fe, Co, and Ni

Sarah J. Watzman[1], Rembert A. Duine[2], Yaroslav Tserkovnyak[3], Stephen R. Boona[1], Hyungyu Jin[4], Arati Prakash[5], Yuanhua Zheng[1], and Joseph P. Heremans[1,5,6]

1. Department of Mechanical and Aerospace Engineering, The Ohio State University, Columbus, Ohio 43210, United States
2. Institute for Theoretical Physics and Center for Extreme Matter and Emergent Phenomena, Utrecht University, Leuvenlaan 4, 3584 CE Utrecht, The Netherlands
3. Department of Physics and Astronomy, University of California, Los Angeles, California 90095, United States
4. Department of Mechanical Engineering, Stanford University, Stanford, California 94305, United States
5. Department of Physics, The Ohio State University, Columbus, Ohio
6. Department of Materials Science Engineering, The Ohio State University, Columbus, Ohio 43210, United States

**Abstract**

Magnon-drag is shown to dominate the thermopower of elemental Fe from 2 to 80 K and of elemental Co from 150 to 600 K; it is also shown to contribute to the thermopower of elemental Ni from 50 to 500 K. Two theoretical models are presented for magnon-drag thermopower. One is a hydrodynamic theory based purely on non-relativistic, Galilean, spin-preserving electron-magnon scattering. The second is based on spin-motive forces, where the thermopower results from the electric current pumped by the dynamic magnetization associated with a magnon heat flux. In spite of their very different microscopic origins, the two give similar predictions for pure metals at low temperature, allowing us to semi-quantitatively explain the observed thermopower of elemental Fe and Co without adjustable parameters. We also find that magnon-drag may contribute to the thermopower of Ni. A spin-mixing model is presented that describes the magnon-drag contribution to the Anomalous Nernst Effect in Fe, again enabling a semi-quantitative match to the experimental data without fitting parameters. Our work suggests that particle non-conserving processes may play an important role in other types of drag phenomena, and also gives a predicative theory for improving metals as thermoelectric materials.

**1. Introduction**

Multi-component fluids and gases are abundant in nature and exist at all scales, ranging from the universe[1] (composed of various types of matter and energy), to cold-atom systems[2] (composed of different types of atoms). Often, the interactions between the various components give rise to new and interesting physics. Examples include the interplay between superfluid and normal components of liquid helium that give rise to second sound,[3] and spin-Coulomb drag[4] that arises due to the interaction between different spin species.

A recent example of a two-component system is magnons that interact with electrons at an interface between a magnetic insulator and a normal metal. This interaction underpins various novel physical effects such as the Spin-Seebeck Effect,[5] spin-Hall magnetoresistance,[6] and the attenuation of magnetization relaxation by electric current through the normal metal. However, magnons also exist in magnetic metals in which they interact with electrons not at interfaces, but throughout the bulk of the material. One measurable consequence of this is the magnon-drag thermopower: the contribution to the thermopower that results from the magnonic heat flux dragging along the electronic charge carriers.

Blatt et al.[7] suggested that magnon-drag might be the dominant mechanism behind the high thermopower ($\alpha$) of elemental iron. Magnon-drag was again suggested as the mechanism underpinning the field-dependence of the thermopower of a Permalloy thermopile,[8] but no proof or quantitative theory was offered in either work. Grannemann and Berger[9] measure an 8% variation in the magnetic field dependence of the Peltier coefficient of a $Ni_{66}Cu_{34}$ at 4 K, which they attribute to a magnon-drag contribution that is gradually destroyed by the magnetic field. Lucassen et al.[10] proposed a contribution to magnon-drag where the magnetization dynamics associated with a thermal flux of magnons pump an electronic spin current due to so-called spin-motive forces. Because of the spin-polarization of the charge carrier, this electronic spin current results in a charge current or voltage.

Here we present a study of magnon-drag that is supported by a basic understanding of the underlying physics. Two theories for the magnon-drag thermopower ($\alpha_{md}$) are presented: a classical hydrodynamic theory based on Galilean-invariant magnon-electron interactions and a theory based on spin-orbit coupling. We outline under what conditions these theories give the same results. We apply them to the thermopower of Fe, Co, and Ni without using any adjustable parameters. The theories are then compared to experimental results for the thermopower of these elemental transition metals.

We further present a semi-quantitative model for the magnon-drag Anomalous Nernst Effect (ANE) based on spin-mixing, and we apply that to the ANE coefficient of single-crystal Fe for which we present the first temperature-dependent data. Measurements of other components of the thermomagnetic tensor, namely the longitudinal and transverse magneto-thermopower and the Planar Nernst Effect (PNE) of single-crystal Fe, are also reported.

Magnon-drag offers a pathway to increase the thermopower and therefore the thermoelectric figure of merit of metals. The models presented here offer the guiding principles for the optimization of metallic thermoelectric alloys, which would have major advantages over the thermoelectric semiconductors used today. Indeed, metals are mechanically stronger than semiconductors, can be formed in net shapes and welded such that thermoelectric elements can be structurally integrated with heat exchangers, and can be heat and corrosion resistant. Beyond its impact on research in thermoelectricity, the more general relevance of our results is the fact that the two contributions to the drag theoretically considered here are very different in nature. One relies on spin-conserving scattering and the other requires spin-flip scattering and/or spin-orbit coupling. Our work thus suggests that processes equivalent to the latter in the context of other drag phenomena give important contributions that have not yet been considered in detail, such as tunneling events in a semiconductor Coulomb drag set-up[11] and electron-phonon spin-flip scattering induced by spin-orbit coupling in electron-phonon drag[12].

## 2. Thermopower

### 2.1 Hydrodynamic theory

In the hydrodynamic theory,[9] the magnons and electrons are modelled as two interpenetrating fluids, and Galilean invariance is assumed such that the electrons and magnons are described by a single parabolic band. Furthermore, Umklapp and magnon non-conserving processes are neglected. As a result, the sign of the thermopower is solely determined by the sign of the charge carriers' effective charge $e$, and thus their effective mass. In what follows, for carriers with a positive effective mass (conduction band electrons), $e < 0$, while $e > 0$ for charge carriers with negative effective mass (valence band holes). The first fluid is the electrons with momentum density $\vec{p}_e = n_e m \vec{v}_e$ in terms of their number density $n_e$, mass $m$, and drift velocity $\vec{v}_e$. The other fluid is composed of magnons with momentum density $\vec{p}_m$, mass $M = \dfrac{\hbar^2}{2D}$ ($D$ is the magnetic exchange stiffness), and $\vec{p}_m = n_m M \vec{v}_m$. The phenomenological equations for the fluid are:

$$\frac{d\vec{v}_e}{dt} = \frac{e}{m}\left(\vec{E} - \alpha_d \vec{\nabla} T - \frac{\vec{v}_e}{\tau_e}\right) - \frac{\vec{v}_e - \vec{v}_m}{\tau_{me}}$$

$$\frac{d\vec{v}_m}{dt} = -\frac{e}{M}\alpha_m \vec{\nabla} T - \frac{\vec{v}_m}{\tau_m} - \frac{\vec{v}_m - \vec{v}_e}{\tau_{em}} \qquad (1)$$

where $\vec{E}$ is the electric field, and $\tau_e$ and $\tau_m$ are transport mean-free times for the electrons and magnons, respectively. The magnonic thermopower is $\alpha_m = \frac{2}{3}\frac{C_m}{n_m e}$, where $C_m$ is the magnon specific heat capacity per unit volume. This is derived by considering magnons as a free, ideal gas with a parabolic dispersion relation, and taking the gradient of the relation $P = \frac{2}{3}U$ between the pressure $P$ and internal energy density $U$ (in units of energy per volume) in the presence of a temperature gradient. The time scales $\tau_{me}$ and $\tau_{em}$ parametrize the magnon-electron collision rate. Thus, according to the conservation of linear momentum, $\frac{n_e m}{\tau_{me}} = \frac{n_m M}{\tau_{em}}$.

Under steady-state conditions and for zero electric current ($v_e = 0$), the previous equations are solved to determine the electric field required to counteract the thermal gradient. The magnon-drag thermopower then becomes:

$$\alpha_{md} \equiv \frac{|\vec{E}|}{|\vec{\nabla}T|} = \frac{2}{3}\frac{C_m}{n_e e}\frac{1}{1+\frac{\tau_{em}}{\tau_m}} \qquad (2).$$

To this one adds the electronic diffusion thermopower $\alpha_d$, given by:[13]

$$\alpha_d = \frac{(\pi k_B)^2 T}{3 e E_F} \qquad (3).$$

Here, $E_F$ is the Fermi energy. The total electron thermopower, including both the diffusive and magnon-drag contributions, but neglecting electron-phonon drag, is:

$$\alpha = \alpha_{md} + \alpha_d \tag{4}$$

In the presence of sufficiently strong disorder scattering, we are allowed to assume at sufficiently low temperatures an energy-independent disorder-dominated magnon mean-free path $l$. Consequently, we expect that $\tau_m$ scales with temperature as $\tau_m^{-1} \propto \sqrt{T}$ because the density of states varies with energy as $\sqrt{\varepsilon}$. At higher temperatures, but low enough to ignore magnon-conserving magnon-phonon interactions, scattering is likely to be dominated by magnon non-conserving processes parameterized by the Gilbert damping parameter $\alpha_{GD}$[14] so that $\tau_m^{-1} \propto \alpha_{GD} T$. The crossover takes place at $T^* \sim \frac{T_c}{s^{2/3}}(\alpha_{GD} l)^2$. Using $\alpha_{GD} \sim 10^{-2}$, $a \sim 1 nm$, and scattering length $l \sim 1$ μm, we obtain $T^* \sim 10^{-2} T_c$ with a range of $T^* \sim (10^{-1}$ to $10^{-3}) T_c$. The electron-magnon scattering frequency is expected to scale with temperature as $\tau_{em}^{-1} \propto T^2$. This results from the combination of momentum and energy conservation constraints for electron-magnon scattering, which give a factor of $\sqrt{T}$, and the reduced phase space for occupied magnon states, which gives a factor of $T^{3/2}$. Thus, $\tau_{em}^{-1}$ decreases with temperature with a higher power of $T$ than $\tau_m^{-1}$, and the factor $\left(1 + \frac{\tau_{em}}{\tau_m}\right)^{-1}$ should vanish as temperature approaches zero. In the limit parameterized by Gilbert damping, the attenuation of the magnon-drag thermopower is expected to have a linear dependence on $T$. Conversely, in the regime where $\tau_m$ is dominated by magnon-phonon scattering, $\tau_m^{-1}$ would vanish faster than $\tau_{em}^{-1}$ due to the rapidly shrinking phase space associated with the linearly-dispersing phonons, and the factor $\left(1 + \frac{\tau_{em}}{\tau_m}\right)^{-1}$ would approach unity. This corresponds to the clean case where magnon-conserving magnon-phonon scattering of momentum is faster than magnon non-conserving processes parameterized by the Gilbert damping, and this is likely to be the case for high-purity elemental metals.

**2.2 Theory based on spin-motive forces**

In addition to the hydrodynamic contribution, we acknowledge a contribution to the magnon-drag thermopower that ultimately stems from spin-orbit interactions.[10] This contribution is parameterized by a dimensionless material parameter $\beta$, typically of the order of 0.01-0.1. It arises from the electric current pumped by the dynamic magnetization associated with a magnon heat flux (via the aforementioned spin-motive forces[15,16]). As was shown in Ref. [10], the electric current density is given by:

$$\vec{j}_e = \sigma\left(\vec{E} + \beta p_s \frac{\hbar}{2e} \frac{\vec{j}_{Q,m}}{sD}\right) \tag{5}$$

where $\sigma = \frac{n_e e^2 \tau_e}{m}$ is the electrical conductivity, $\tau_e$ is the electronic transport relaxation time, $p_s$ the spin polarization of the electric current (typically of order 1), $\vec{j}_{Q,m}$ is the magnon heat current, and $D$ is the spin stiffness. Combining Eq. (5) with Fourier's law for magnons, $\vec{j}_{Q.m} = -\kappa_m \vec{\nabla} T$, and assuming diffusive magnon transport and a boundary condition of an electrically open circuit in the sample, leads to:

$$\alpha'_{md} = \beta p_s \frac{\hbar}{2e} \frac{\kappa_m}{sD} \tag{6}.$$

In the simplest microscopic models,[17, 18, 19] the sign of this thermopower also depends on the sign of the effective mass. Based on Landau-Lifschitz-Gilbert phenomenology, Flebus et al. [20] pointed out a Berry-phase correction to the above result which amounts to replacing $\beta$ with $\beta^* = \beta - 3\alpha_{GD}$ in Eq.(6). In our discussion and in Eq. (6), therefore, the effective $\beta$ entering the expressions for the magnon-drag thermopower should be understood as $\beta^*$. We note in passing that this factor can also affect the sign of the magnon-drag Seebeck coefficient (depending on the ratio of $\beta/\alpha_{GD}$ which can be estimated[21] to be of the order of 1 to 10), irrespective of the effective mass considerations.

To compare Eq. (2) and Eq. (6), we estimate $\kappa_m$ and $C_m$. Assuming that the magnon dispersion is quadratic (i.e. at sufficiently low magnon energies and $T < T_c$), $C_m \sim k_B s \left(\dfrac{T}{T_c}\right)^{3/2}$, where $s \sim a^{-3}$ (in units of $\hbar$) is the saturation spin density, $T$ is the temperature, $k_B$ is the Boltzmann constant, and $T_c$ is the Curie temperature of the ferromagnet, the hydrodynamic formula for magnon-drag thermopower (Eq. 2) can be rewritten:

$$\alpha_{md} \approx \frac{k_B}{e} \frac{s}{n_e} \left(\frac{T}{T_c}\right)^{3/2} \frac{1}{1+\dfrac{\tau_{em}}{\tau_m}} \qquad (7).$$

To estimate the thermopower due to spin-motive forces, Eq. (6), we take $\kappa_m \sim k_B^2 T \left(\dfrac{T}{T_c}\right) s^{2/3} \dfrac{l}{\hbar}$ at $T<T^*$ and $\kappa_m \sim k_B^2 T \sqrt{\dfrac{T}{T_c}} s^{1/3} \Big/ \alpha_{GD} \hbar$ at $T>T^*$ due to magnon diffusion.[22] Using the latter expression and $k_B T_c \sim s^{2/3} D$, we obtain:

$$\alpha'_{md} \sim \frac{\beta p_s}{\alpha_{GD}} \frac{k_B}{e} \left(\frac{T}{T_c}\right)^{3/2} \qquad (8).$$

The two magnon-drag contributions to thermopower stem from different microscopic physics: the hydrodynamic contribution is nonrelativistic and the contribution due to spin-motive forces is based on spin-orbit interactions that are intrinsically relativistic and non-hydrodynamic as they do not conserve magnons. Remarkably, the contributions estimated in Eqs. (7) and (8) yield comparable values if we set $\frac{s}{n_e} \sim 1$ and $\frac{\beta p_s}{\alpha_{GD}} \sim 1$, which are certainly reasonable values for pure elemental transition metals, and omit the last, scattering time-dependent factor in Eq. (7). However, there are various transport regimes in which the hydrodynamic contribution and spin-motive force contribution to magnon-drag thermopowers are clearly distinct.

**2.3    Experimental thermopower data**

We present experimental data for the thermopower $\alpha$ of various samples of Fe, Co, and Ni in Fig. 1. Fe has an Ordinary Hall Effect of polarity opposite to that of Co and Ni,[23] which reflects the polarity of the effective mass of the dominant charge carriers and the sign of $\alpha$.

The thermopower of Fe has been measured by numerous authors[7,24] over the temperature range included here, but we repeated the measurements because they are extended to the thermomagnetic tensor in the next section. All measurements on Fe were completed on a 7.13 mm × 5.05 mm × 1.07 mm 99.994% pure single-crystal of [100] Fe from Princeton Scientific (red points in Fig. 1). We also took thermopower data on a dense sintered sample of polycrystalline Fe (blue points in Fig. 1a), which gave essentially the same results (see insert in Fig. 1a). The polycrystalline sample was prepared from 99.998% Fe powder in a Spark Plasma Sintering system under a uniaxial 50 MPa pressure, using a 3 minute ramp to 750 °C, a 2 minute hold at 750 °C, then an uncontrolled cooling to room temperature; the sample was at least 94% dense. Two similarly sized 99.9% pure polycrystalline ingots of Co from Alfa Aesar were used for bulk Co measurements, giving consistent results (red points in Fig. 1b). A 50 % porous bulk polycrystalline sample of Co was prepared from 1 g of powder, which was obtained from Alfa Aesar and rated as –22 mesh particle size with 99.998% purity. The powder was placed in a 10 mm diameter graphite die and compacted via spark plasma sintering at 250 °C for 30 minutes under 50 MPa of uniaxial pressure. The resulting pellet was mechanically stable but brittle enough to be cut easily with a handheld wire cutter. Based on the cylindrical pellet's size (2.97 mm thick corresponding to a volume of 0.23 cc) and the amount of Co used (1 g corresponding to a volume of 0.11 cc), we estimate the sample to be ~50% dense. The thermopower of the porous Co sample is reported as blue points in Fig. 1b. The thermopower data on a Ni ingot (red points in Fig. 1c) are taken from the literature.[25] We also prepared a 50 % porous sample from Ni powder following the protocol used for Co, and the thermopower data on this sample are given as blue points in Fig. 1c.

The Thermal Transport Option (TTO) on a 7T and 9T Quantum Design Physical Property Measurement System (PPMS) with customized controls programmed in LabVIEW was used for material characterization between 1.8 K and 400 K. For measurement of the thermopower, a copper heat sink was attached to one end of the sample using silver epoxy. A gold-plated copper plate was attached to the opposite end of the sample using silver epoxy, and a thin lead of the same material was left protruding from the plate for later attachment of the resistive heater assembly. Gold-plated copper leads of width 0.65 mm were attached to the sample using silver epoxy along the sample edge between the heater and heat sink spaced approximately 4 mm apart. The heat sink was clamped to the TTO puck, and gold-plated copper assemblies purchased from Quantum Design containing calibrated Cernox$^{TM}$ thermometers and voltage measurement wires were clamped to the leads. A resistive heater assembly was clamped to the lead on the heater side. Thermopower and resistivity data between 400 K and 1000 K were taken on the Co ingot using a Linseis LSR-3. A sample with a cross section of 2.70 × 4.21 mm was placed in the chamber, and two type S thermocouples were attached to the sample approximately 3.75 mm apart from one another. The chamber was purged with helium gas. Temperature-dependent steady-state measurements of thermal conductivity and resistivity of the single-crystal and sintered samples of Fe, the ingot and porous sample of Co, and the porous sample of Ni are reported in Fig. 2.

The thermopower of Fe is in good agreement with previous work.[7,24] The thermopower of the Co ingot agrees with previous measurements above 150 K,[26] but at lower temperatures it shows a sign reversal near 100 K and a pronounced maximum between 11 and 14 K (we are unaware of previously existing data below 90 K). The porous Co sample does not show either of these features, and this sample displays a negative thermopower that follows closely to a $T^{3/2}$ law up to 400 K. We attribute the positive peak around 12 K in the Co ingot to phonon-drag since it is present in the ingot but not in the porous sample. The Umklapp-limited phonon mean free path in elemental Co is expected to be longer than the grains in the porous sample around 12 K. Therefore boundary scattering will limit the phonon mean-free path in the porous sample, and suppress phonon drag. The thermopower of the Ni ingot,[25] like that of Co, has an additional feature around 20 K that is not present in our porous sample and is attributed to phonon-drag by analogy with the case of Co. Therefore, if magnon-drag lies at the origin of the thermopower of Co or Ni, Eq. (4) is to be tested against the results on the porous samples. No similar additional structure is observed in the thermopower of Fe, which is quite robust vis-à-vis disorder, as was already reported by Blatt et al.[7]

## 2.4   Comparison between theory and experiment

The data on the thermopower of Fe, Co, and Ni are compared to Eqs. (2-4) ($\alpha_{md}$ dashed line and $\alpha$ full line in Fig. 1) in the limit where we assume that electron-magnon scattering dominates all magnon scattering, allowing us to ignore the scattering time-dependent prefactor ($\tau_{em} < \tau_m$). The following numerical values are used. $C_m$ is derived[27] from the experimental magnon dispersion relation for Fe,[28] Co,[29] and Ni[30]. Below energies of about 4 meV, the magnon dispersions are approximately quadratic ($D \approx 2.7 \times 10^{-22}$ eV-m² for Fe, $4.3 \times 10^{-22}$ eV-m² for Co, and $5.9 \times 10^{-22}$ eV-m² for Ni), which leads to Eq. 7. At higher energies (the case for Fe), we calculate $C_m$ from the polynomial fit to the dispersion.[28] The total charge carrier concentrations are[31] $1.7 \times 10^{23}$ cm⁻³ for Fe, $8.9 \times 10^{22}$ cm⁻³ for Co, and $9.2 \times 10^{22}$ cm⁻³ for Ni. We assume that only the s- and p-electrons contribute to transport[13] and derive their concentration from the density of states at the Fermi energy:[32] $n_e \approx n_{sp} = 2.36 \times 10^{21}$ cm⁻³ (Fe), $8.1 \times 10^{21}$ cm⁻³ (Co), and $3 \times 10^{21}$ cm⁻³ (Ni). The sign of the thermopower is derived from the slope of the s- and p-bands' density of states at $E_F$: $\alpha$ is positive for Fe and negative for Co and Ni. Eq. 3 then gives the dashed line representing $\alpha_{md}$ in Fig. 1. The diffusion thermopower in Eq. 3 can be estimated roughly using the Fermi energy ($E_F = 0.48$ eV for Fe, 0.76 eV for Co, 1.9 eV for Ni) for s- and p-electron bands.[32] The band structure involves several pockets of electrons with dominantly s-p and d-character so that these simplified band structure parameters are affected by uncertainties of about a factor of 2. No adjustable parameter is used to fit the lines in Fig. 1. We submit that the similarities in magnitude and temperature dependence of $\alpha$ observed at $T < 80$ K for Fe and $T < 400$ K for the porous sample of Co are evidence that both models are reasonable. Above those temperatures, the data points fall under the calculated line. This could be due to the increased effect of the factor $\left(1+\dfrac{\tau_{em}}{\tau_m}\right)^{-1}$, to the contribution of the additional term $\beta^* = \beta - 3\alpha_{GD}$ discussed above, or to a breakdown of the approximation for the magnon-heat conductivity that was used to estimate the contribution due to spin-motive forces. In addition, the data on the Co ingot show a discontinuity at the face-centered cubic/hexagonal phase transition near 700 K, as reported previously.[26] The data on the porous Ni sample do not agree as well with this simple model: the calculated values are about two times larger than the experimental data on the porous sample.

3. **Thermomagnetic Effects**

3.1 **Spin-mixing theory for the Anomalous Nernst Effect**

*A priori*, neither model for $\alpha_{md}$ presented above accounts for the generation of a skew force. Electric fields perpendicular to the direction of an applied temperature gradient in the presence of an applied magnetic field in the third perpendicular direction can arise from two other mechanisms. First, in ferromagnetic metals that also have strong spin-orbit interactions and a measureable spin-Hall coefficient, one expects a bulk Spin-Seebeck Effect-like contribution to the Nernst coefficient.[33] We are not aware of measurements of the spin-Hall angle of Fe. An interpolation of the spin-Hall angle measurements in 3d elements[34] as a function of their atomic number suggests that Fe has a small spin-Hall angle; therefore, we neglect this contribution. The second possible mechanism for a magnon-drag contribution to the Nernst coefficient arises from spin-mixing, which was suggested for the resistivity[13] and thermopower.[35] This model, inspired by a similar model for the phonon-drag contribution to the Nernst Effect,[36] is presented here.

Consider two independent spin-up and spin-down conduction electron channels with densities ($n\uparrow$ and $n\downarrow$) at $E_F$, partial conductivities ($\sigma\uparrow$ and $\sigma\downarrow$), mobilities ($\mu\uparrow$ and $\mu\downarrow$), Hall coefficients ($R_H\uparrow$ and $R_H\downarrow$), thermopowers ($\alpha\uparrow$ and $\alpha\downarrow$), and Nernst coefficients ($N\uparrow$ and $N\downarrow$). The total Nernst coefficient is then derived by writing the Onsager relation for each channel, adding the fluxes, and solving the proper boundary relations for the transport coefficients, as is done for multi-carrier transport in semiconductors:[37]

$$N = \frac{\left(N_\uparrow \sigma_\uparrow + N_\downarrow \sigma_\downarrow\right)(\sigma_\uparrow + \sigma_\downarrow) + \left(N_\uparrow R_{H\downarrow} + N_\downarrow R_{H\uparrow}\right)\sigma_\uparrow^2 \sigma_\downarrow^2 \left(R_{H\uparrow} + R_{H\downarrow}\right) B^2 + \sigma_\uparrow \sigma_\downarrow \left(\alpha_\uparrow - \alpha_\downarrow\right)\left(\sigma_\uparrow R_{H\uparrow} - \sigma_\downarrow R_{H\downarrow}\right)}{(\sigma_\uparrow + \sigma_\downarrow)^2 + \sigma_\uparrow^2 \sigma_\downarrow^2 \left(R_{H\uparrow} + R_{H\downarrow}\right)^2 B^2} \quad (9)$$

with $B=\mu_0 H_{applied}$ in the field range where Fe's magnetization is saturated. The terms in $B^2$ are neglected because the mobility is low. In principle, $\alpha$ also needs to be considered in light of the two-channel model, but $N$ is more sensitive to this model than $\alpha$, because it is sensitive to the difference ($\alpha\uparrow - \alpha\downarrow$), while $\alpha$ is the conductivity-weighted average between $\alpha\uparrow$ and $\alpha\downarrow$. Considering each channel separately first, a net Lorentz force arises only if electrons have a velocity distribution that does not average out when integrated over the thermal energy spread of several $k_B T$ centered around $E_F$, as happens when there is an energy dependence to the scattering mechanism. Each channel then develops a partial Nernst coefficient, $N\uparrow$ and $N\downarrow$. These result in a Mott-type relation for the Nernst coefficient, relating it in the same way to the Hall coefficient as $\alpha$ is related to $\sigma$. The thermopower of liquid Fe[38] is a function almost exclusively of the energy dependence of the scattering (mostly s-d scattering), and shows experimentally that this contribution is small. Therefore, we assume that $N\uparrow \approx N\downarrow \approx 0$ in Eq. 9. Since $\alpha\uparrow \neq \alpha\downarrow$ and $\mu\uparrow \neq \mu\downarrow$, the term in ($\alpha\uparrow - \alpha\downarrow$) in Eq. 9 becomes dominant:

$$N = \frac{\sigma_\uparrow \sigma_\downarrow (\alpha_\uparrow - \alpha_\downarrow)(\mu_\uparrow - \mu_\downarrow)}{(\sigma_\uparrow + \sigma_\downarrow)^2} \qquad (10).$$

We assume further that $\alpha\uparrow$ and $\alpha\downarrow$ are $\alpha_{md}\uparrow$ and $\alpha_{md}\downarrow$ (Eq. 2), and that $n\uparrow \approx n_{sp}\uparrow$ and $n\downarrow \approx n_{sp}\downarrow$. These are proportional to the density of states at the Fermi level, i.e. $n_{sp\uparrow\downarrow} \propto \mathcal{D}_{sp\uparrow\downarrow}$, which are known.[32] The partial conductivities for each channel, up or down ($\uparrow\downarrow$), are $\sigma_{\uparrow\downarrow} = n_{sp\uparrow\downarrow} e \mu_{\uparrow\downarrow}$ with mobilities $\mu_{\uparrow\downarrow} = \mu_{sp\uparrow\downarrow} = e \frac{\tau_{sp\uparrow\downarrow}}{m_{sp\uparrow\downarrow}}$ given as a function of scattering frequencies and effective masses. The partial thermopowers are $\alpha_{\uparrow\downarrow} \approx \alpha_{md\uparrow\downarrow} = \dfrac{2}{3}\dfrac{C_m}{n_{sp\uparrow\downarrow}e}$, and Eq. 10 can be expressed in terms of the ratios $r_n \equiv \dfrac{n_{sp\downarrow}}{n_{sp\uparrow}}$, $r_\mu \equiv \dfrac{\mu_{sp\downarrow}}{\mu_{sp\uparrow}}$. Thus, we can further reduce Eq. 10 to the following:

$$N = \frac{2}{3}\frac{C_m}{\rho n_{sp}e^2}\frac{(1-r_n^{-1})(1-r_\mu)r_n r_\mu}{(1+r_n r_\mu)^3} \tag{11}$$

where $n_{sp} = n_{sp\uparrow} + n_{sp\downarrow}$, and the carrier mobility is derived from the sample's resistivity $\rho$.

To estimate the mobility ratios, we take the effective masses as proportional to the density of states at the Fermi level to the 2/3 power, $m_{sp\uparrow\downarrow} \propto \mathcal{D}_{sp\uparrow\downarrow}^{2/3}$. Assuming that s-d scattering dominates, we further assume that this mechanism is spin-selective, i.e. that the scattering frequency of electrons in the spin-up channel is proportional to the density of states of spin-up d-electron bands: $\tau_{sp\uparrow\downarrow}^{-1} \propto \mathcal{D}_{d\uparrow\downarrow}^{-1}$, so that $\mu_{sp\uparrow\downarrow} \propto \mathcal{D}_{d\uparrow\downarrow}\mathcal{D}_{sp\uparrow\downarrow}^{-2/3}$. With the band parameters of Ref. [32], Eq. 11 becomes:

$$N \approx 0.05\frac{C_m}{\rho n_{sp}e^2} \approx 0.07\frac{\alpha_{md}}{\rho} \tag{12}.$$

In the low-temperature limit, $\rho \propto T^0$, $\alpha_{md} \propto C_m \propto T^{3/2}$, and Eq. (12) predicts that at low temperature $N \propto T^{3/2}$. Eq. (12) is compared to experimental data in the following section.

**3.2  Experimental Nernst thermopower and Anomalous Nernst Effect coefficient of Fe**

The components of the thermomagnetic transport tensor in a magnetic field are denoted $\alpha_{ABC}$, where $A$ designates the direction of the applied heat flux and temperature gradient, $B$ designates the direction of the measured electric field, and $C$ designates the direction of the applied magnetic field, i.e. $\alpha_{ABC} \equiv E_B / \nabla T_A \big|_{H_C}$. The third index is generally omitted for the thermopower at zero field, but since we are describing bcc-Fe, polycrystalline Co, and polycrystalline Ni, where the thermopower is isotropic, we omitted all subscripts in the first section and denoted the thermopower at zero field as simply $\alpha$. The magneto-thermopower in a longitudinal magnetic field is then $\alpha_{xxx}$ and in a transverse field $\alpha_{xxz}$. The Nernst thermopower is $\alpha_{xyz}$; the planar Nernst thermopower is $\alpha_{xyx}$.

To the best of our knowledge, prior to the data presented here, only values for the Nernst coefficient near room temperature are reported in the literature, but no systematic data as a function of field and temperature have been published yet. Historical references are by Zahn,[39] Hall,[40] Butler,[41] and Smith[42]. The single-crystal Fe sample is aligned such that $x$ is the [100] axis, $y$ along [010], and $z$ along [001]. The measurements were carried out in the TTO system described above using the static heater-and-sink method.[43] A third gold-plated copper lead was added to the sample (also attached using silver epoxy), mounted as stated previously, directly opposite of the lead closest to the heat sink. The voltage wire from the heater-side of the Cernox™ assembly was removed and soldered to the new lead. The sample was rotated such that the magnetic field was applied in the appropriate direction. Data was taken at discrete temperatures ranging from 1.8 K to 400 K with magnetic fields sweeping in both directions between -90 kOe and 90 kOe at multiple magnetic field ramp rates.

The Nernst thermopower $\alpha_{xyz}(H_{a,z})$ of Fe is shown as a function of magnetic field in Fig. 3 in two field ranges. When plotted from -90 to +90 kOe, the Anomalous Nernst Effect (ANE) is clearly in evidence from -20 to 20 kOe, and the Ordinary Nernst Effect (ONE) outside this range. The bottom frame zooms in on the ANE field range, where hysteresis is observed from -5 to 5 kOe: this is attributed to the motion of the direction of magnetization in domains inside the sample and will be discussed later in the context of the Planar Nernst Effect (PNE). The Nernst coefficient $N \equiv \dfrac{\partial \alpha_{xy}(H_{a,z})}{\partial H_{a,z}}$ derived in the ANE regime is shown as a function of temperature in Fig. 4. The temperature dependence of the ANE slope $N$ follows the $T^{3/2}$ law discussed above, suggesting a magnonic origin.

A semi-quantitative comparison of the data in Fig. 4 with Eq. (12), using the experimental values for $\alpha$ (Fig. 1) and $\rho$ (Fig. 2), gives the solid curve in Fig. 4. This procedure again uses no adjustable parameters. The agreement with the data is reasonable up to about 200 K, a higher temperature than for the thermopower, which is expected since the experimental values of thermopower are used in Eq. 12 to obtain the solid curve. Above this temperature, the experimental data continue to increase with temperature while Eq. 12 saturates. The fit can be improved by adding a negative constant contribution to $N$ of -50 nV K$^{-1}$T$^{-1}$.

### 3.3 Longitudinal and transverse magneto-thermopower

To the best of our knowledge, besides the zero-field thermopower of Fe,[7,24] only experimental results on $\alpha_{xxx}$ and $\alpha_{xxz}$ of Fe at 0.2 T are reported in the literature.[44] Measurements of both the temperature dependence of the longitudinal ($\alpha_{xxx}$) and transverse ($\alpha_{xxz}$) magneto-thermopower were completed here. The values for $\alpha_{xxx}(|H_{a,x}| \leq 70$ kOe) do not deviate measurably from $\alpha$: no longitudinal magneto-thermopower effect is resolved above the error bar of the present measurements, which is limited by the noise floor of 50 nV to about 0.2% on relative measurements. Blatt et al.'s in-field data[44] are internally inconsistent, since Fig. 3 in Ref [44] shows no magnetic field dependence to $\alpha_{xxx}$, consistent with our observations, but Figs. 1 and 2 in Ref [44] show a difference between $\alpha_{xxx}(H_{a,x} = 2 \text{ kOe})$ and $\alpha_{xxx}(H_{a,x} = 20 \text{kOe}) = \alpha$, which is not reproduced here.

The transverse magneto-thermopower $\alpha_{xxz}(H_{a,z})$ is reported as relative values for the change of $\alpha_{xxz}(H_{a,z})$ vis-à-vis $\alpha_{xxz}(H_{a,z}=0 \text{ kOe})$ in Fig. 5 as a function of $H_{a,z}$ at various temperatures. The relative effect is a small increase in thermopower, which is not resolved below 100 K. In principle, an applied magnetic field opens an energy gap $g\mu_B H$ in the magnon spectrum of ferromagnets (here, $\mu_B$ is the Bohr magneton, and $g$ is the Landé factor, which is about 2 for Fe). In practice, this gap is too small at 70 kOe to have a resolvable effect on $C_m$ and $\alpha_{md}$ above about 10 K, given the accuracy of our measurements. Below 10 K, the thermopower is still dominated by electronic diffusion, and the magnitude of the magnon-drag contribution to the thermopower is too small to resolve its magnetic field dependence. Therefore, the most likely cause for the magneto-thermopower effect in $\alpha_{xxz}$ is not related to changes in magnon density, but perhaps due to the spin-mixing effects, which were not taken into account during the calculations of the net thermopower.

### 3.4 Planar Nernst Effect

For the measurements in the $\alpha_{xyx}$ and $\alpha_{xyy}$ PNE geometries, the sample was mounted for the TTO system in the same manner as the Nernst geometry but rotated to apply the magnetic field in the appropriate directions. Data were taken at discrete temperatures ranging from 1.8 K to 400 K with magnetic fields sweeping in both directions between -70 kOe and 70 kOe at multiple magnetic field ramp rates.

A non-zero planar Nernst thermopower $\alpha_{xyx}$, exceeding the noise level of 50 nV, is shown in Fig. 6 as a function of magnetic field. The signal is an even function of magnetic field and saturates around the field value where the magnetization of the sample saturates. The difference between the zero field value (set to be zero) and the saturation value is plotted as a function of temperature in Fig. 7. This value increases rapidly with decreasing temperature below 50 K, but it is non-zero and nearly constant between 50 and 300 K. No signal is detected for $\alpha_{xyy}$ except for noise transients at what amounts to the coercive field of the sample in that geometry, such that, for all practical purpose, $\alpha_{xyy} \approx 0$ in our measurements. This is consistent with the observations of Pu et al.[45] The PNE is associated with the switching of the magnetization of the sample.[45] The magnetic field range over which a signal change is observed in $\alpha_{xyx}$ does correspond to the field range over which hysteresis is observed in $\alpha_{xyz}$ (Fig. 3) and is only a fraction of the extent of the ANE field range; *a posteriori*, this justifies attributing that feature in the ANE to the switching of a domain in the sample.

## 4. Conclusion

In conclusion, this paper describes both hydrodynamic and relativistic contributions to the magnon-drag thermopower and a spin-mixing model for the magnon-drag Nernst coefficient at magnetic fields above the saturation magnetization. We have shown that the thermopower theories can, depending on which scattering processes limit the electronic and magnon transport, coincide at low temperatures. The theories explain the experimental thermopower of Fe and Co, which have two different polarities, semi-quantitatively without adjustable parameters. The results are less conclusive about the thermopower of Ni. The theories presented also have predictive power, potentially enabling the design of metallic thermoelectric alloys that might become competitive with semiconductor thermoelectrics. For example, Eq. 2 shows that alloys with a lower concentration of s- and p-electrons than those of elemental Fe, Co, and Ni are expected to have a higher $\alpha_{md}$ and therefore figure of merit ($ZT = \frac{\alpha^2 \sigma}{\kappa} T$). Note that such optimization does not require changing the overall concentration of electrons in a metal, which would be a daunting task, but involves the redistribution of free electrons between bands of s-p orbital character and bands of d-orbital character. The contribution due to spin-motive forces may be increased by increasing the ratio of $\beta$ to the Gilbert damping, as has been achieved, for example, by alloying Permalloy with vanadium.[46] A further possibility is tuning the ratio $\beta/\alpha_{GD}$ in composite materials by introducing second phases with the presence of interfaces that affect the Gilbert damping and $\beta$-parameter differently.[47]

**Acknowledgements**

We acknowledge support from the U. S. Army Research Office MURI grant number W911NF-14-1-0016. SJW is supported by the U. S. National Science Foundation Graduate Research Fellowship Program under Grant No. DGE-0822215 and YZ by a fellowship from The Ohio State University. RD is supported by the Stichting voor Fundamenteel Onder zoek der Materie (FOM) and is part of the D-ITP consortium, a program of the Netherlands Organisation for Scientific Research (NWO) that is funded by the Dutch Ministry of Education, Culture and Science (OCW).

**Figure Captions**

Figure 1. Temperature dependence of the thermopower of Fe (a), Co (b), and Ni (c). The insets represent data on a linear scale; the main frames on a logarithmic scale (negative for Co and Ni since they have a negative thermopower). The thermopower is given on two Fe (a) samples, a 95%-dense sintered polycrystal (red triangles) and a single crystal (black circles) with the heat

flux oriented along the <100> axis. The thermopower is given on two Co (b) and two Ni (c) samples, polycrystalline ingots (red triangles) and 50% porous samples (blue squares) prepared to eliminate the effects of phonon-drag. The data for the Ni ingot are taken from the literature.[25] The dashed black lines give the magnon-drag thermopower calculated from Eq. (2) with $\tau_m/(1+\tau_{em})=1$, as explained in the text; the full black lines are the sum of the magnon-drag and diffusion thermopower, Eq. (4). The agreement for Fe is excellent, and is within about 40% for Co. The thermopower of Ni is about two times smaller than the calculation suggests.

Figure 2. Temperature dependence of electrical resistivity $\rho$ (a) and thermal conductivity $\kappa$ (b) of the samples whose thermopower is reported in Fig. 1. The data on Fe are shown as full circles (black for the single crystal, red for the 95% dense polycrystals). The data on Co are given as open squares, and on Ni as closed triangles, red for the polycrystalline ingots, and blue for the porous samples.

Figure 3. Dependence of the Nernst thermopower $\alpha_{xyz}$ on an applied external magnetic field $H_{a,z}$ for single crystal Fe, with $x$ // <100> axis. Frames (a) and (b) give magnetic field dependencies at the temperatures indicated over two different field ranges. The Anomalous Nernst Effect (ANE) extends from about -20 to 20 kOe, the Ordinary Nernst Effect (ONE) outside this range. Hysteresis is visible in the inner loop in frame (b) and is likely due to domain realignments. The inset shows the geometry of the measurements.

Figure 4. Temperature dependence of the anomalous Nernst coefficient $N_{xyz} \equiv \partial \alpha_{xyz} / \partial H_{a,z}$, which is the slope of the Nernst thermopower in the ANE regime. The full line represents the model calculation of Eq. 12.

Figure 5. Magnetic field dependence of the transverse magneto-thermopower $\alpha_{xxz}$, normalized to the zero field thermopower. The inset shows the geometry of the measurements with $x$ // <100> axis.

Figure 6. The magnetic field dependence of the planar Nernst thermopower $\alpha_{xyx}$ at selected temperatures. The inset shows the geometry of the measurements, with $x$ // <100> axis.

Figure 7. The temperature dependence of the planar Nernst thermopower $\alpha_{xyx}$ in the field range where it is saturated ($H_{a,x} > 5$ kOe). Error bars represent a 97% confidence interval for the standard error.

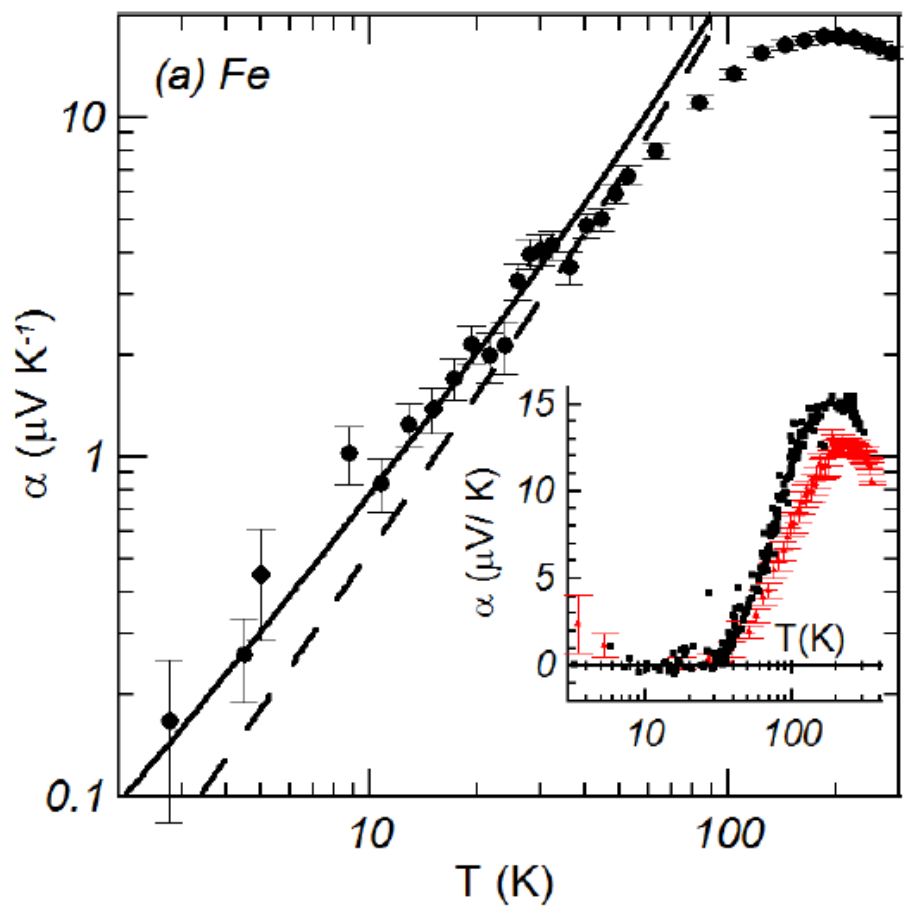

Fig. 1a

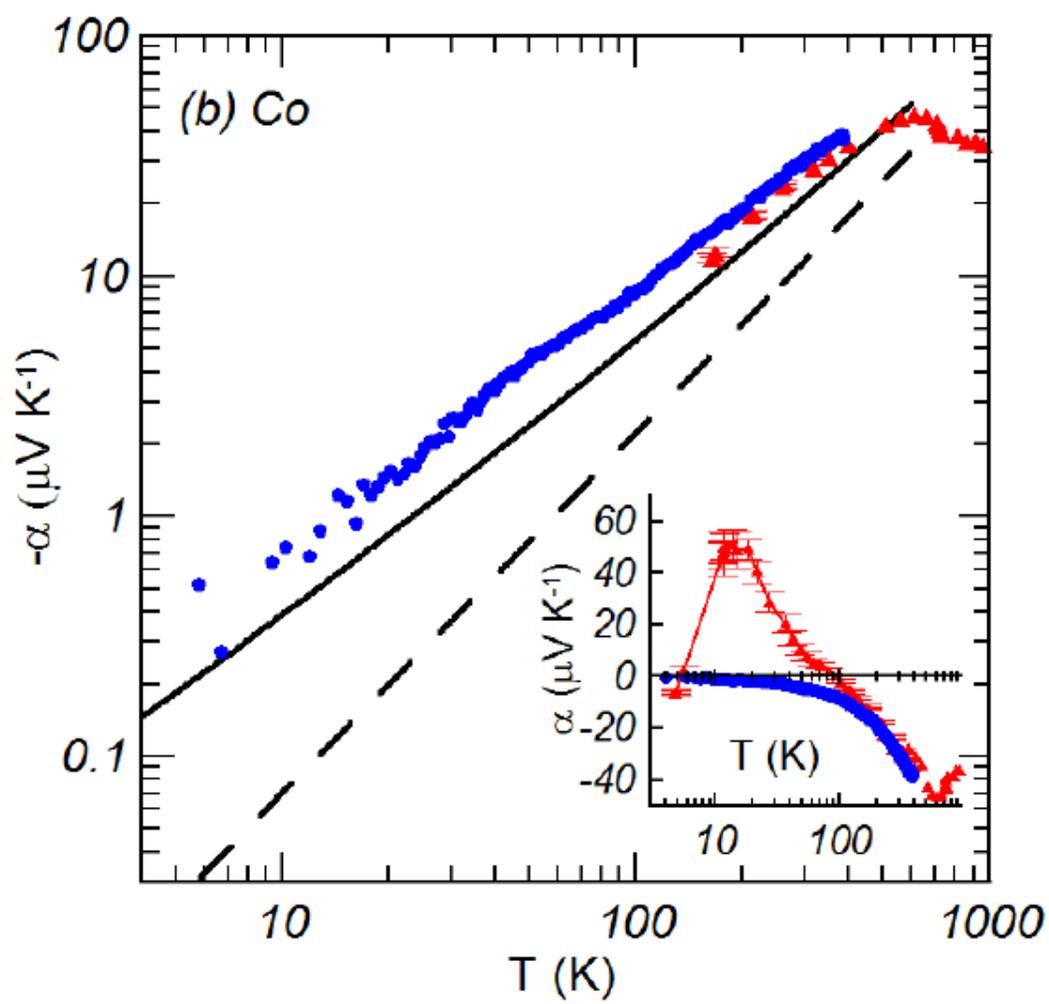

Fig. 1b

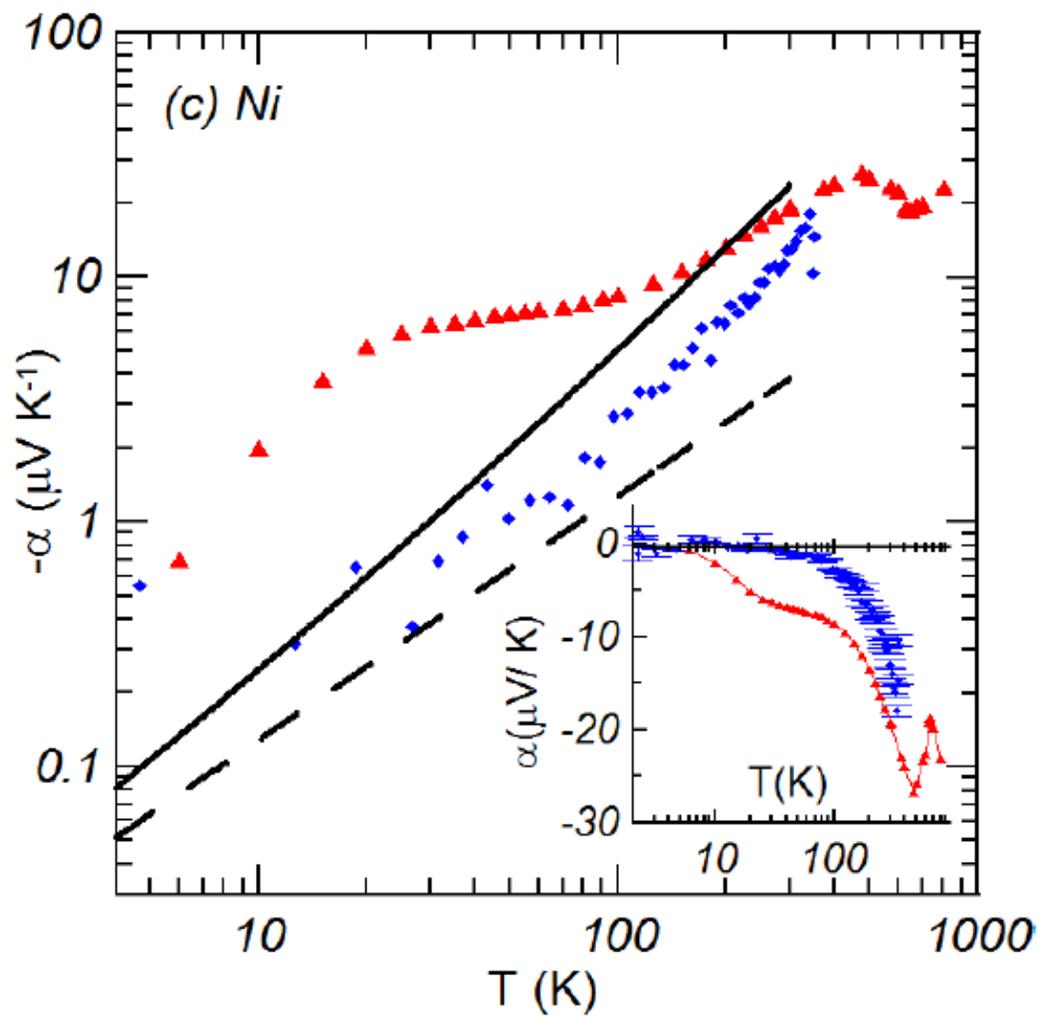

Fig. 1c

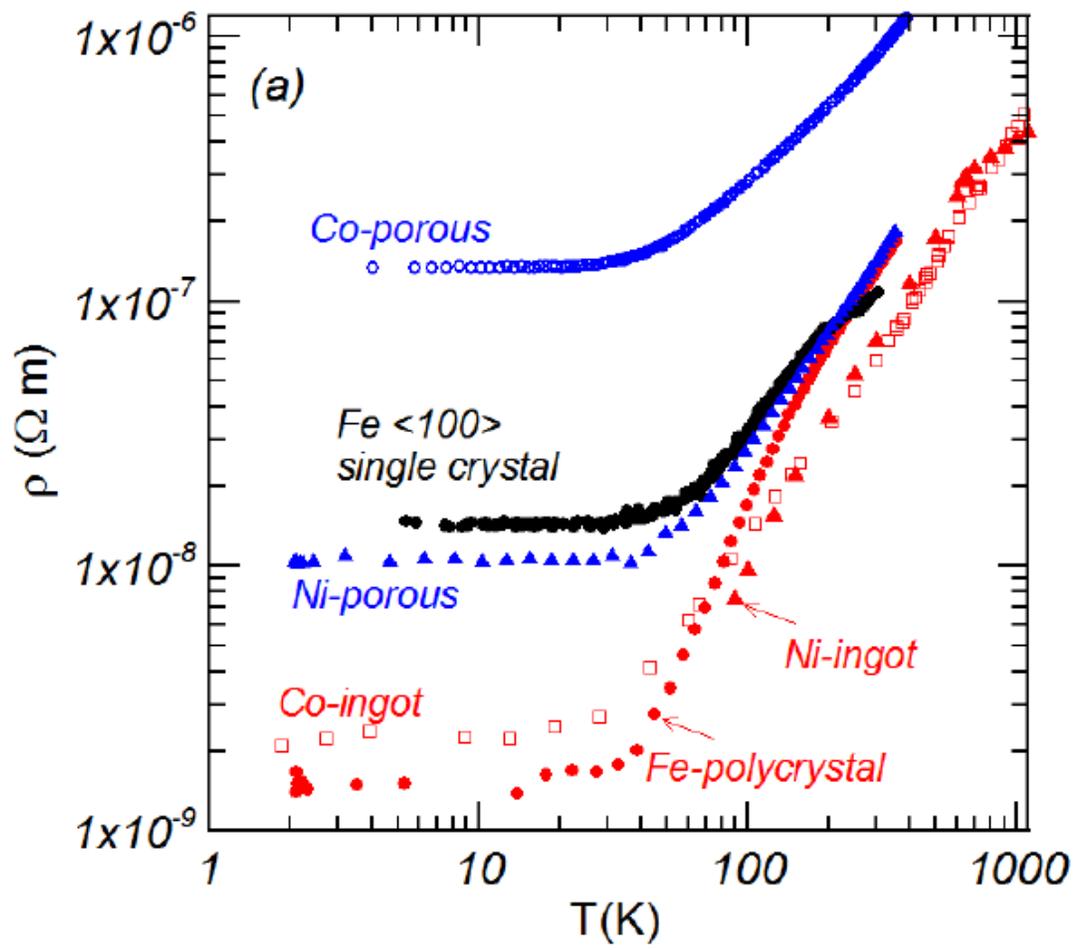

Fig. 2a

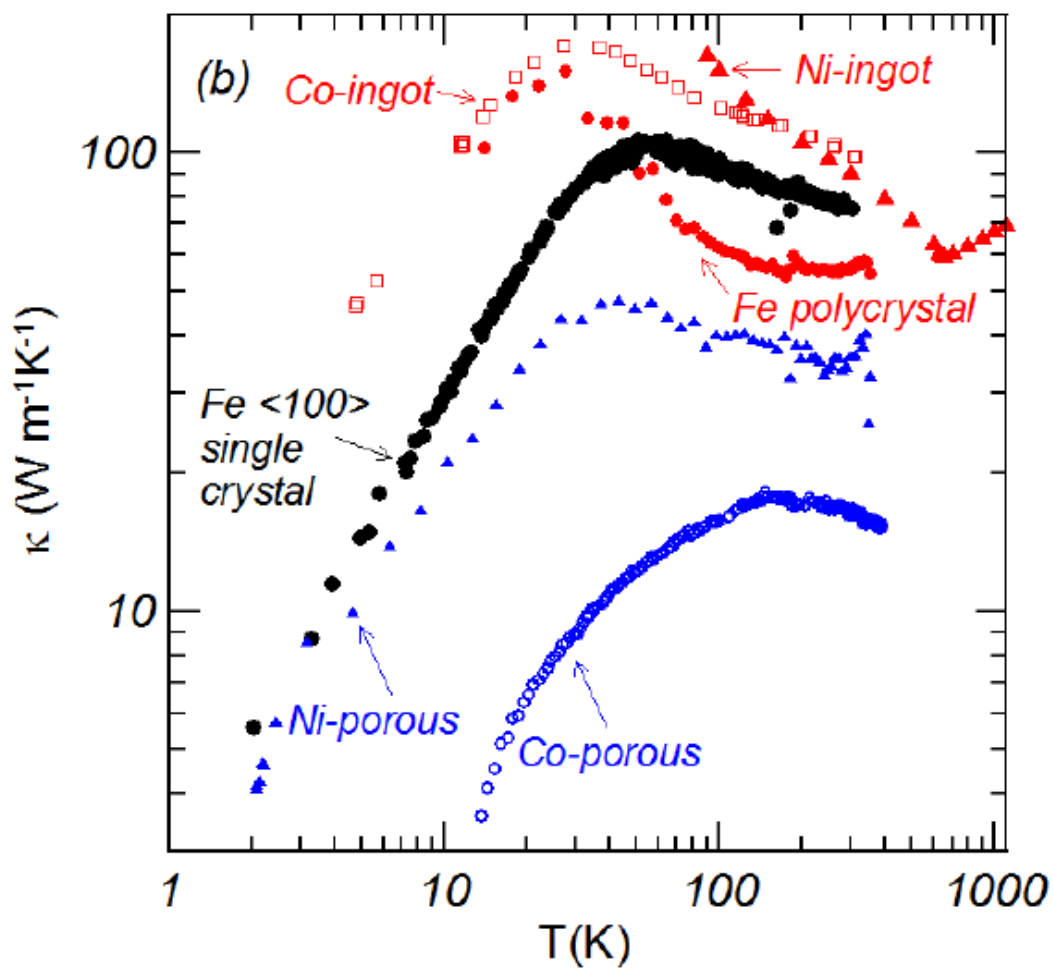

Fig. 2b

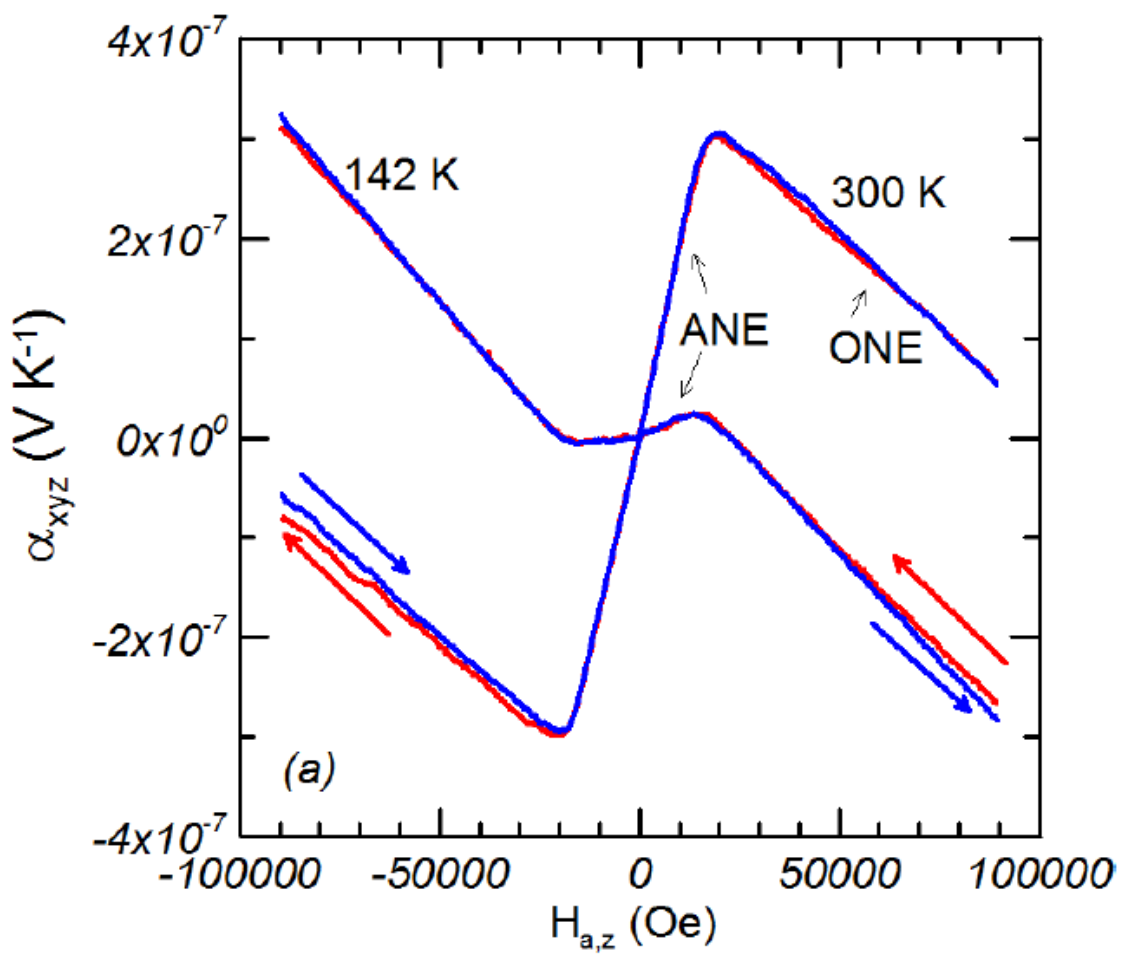

Fig. 3a

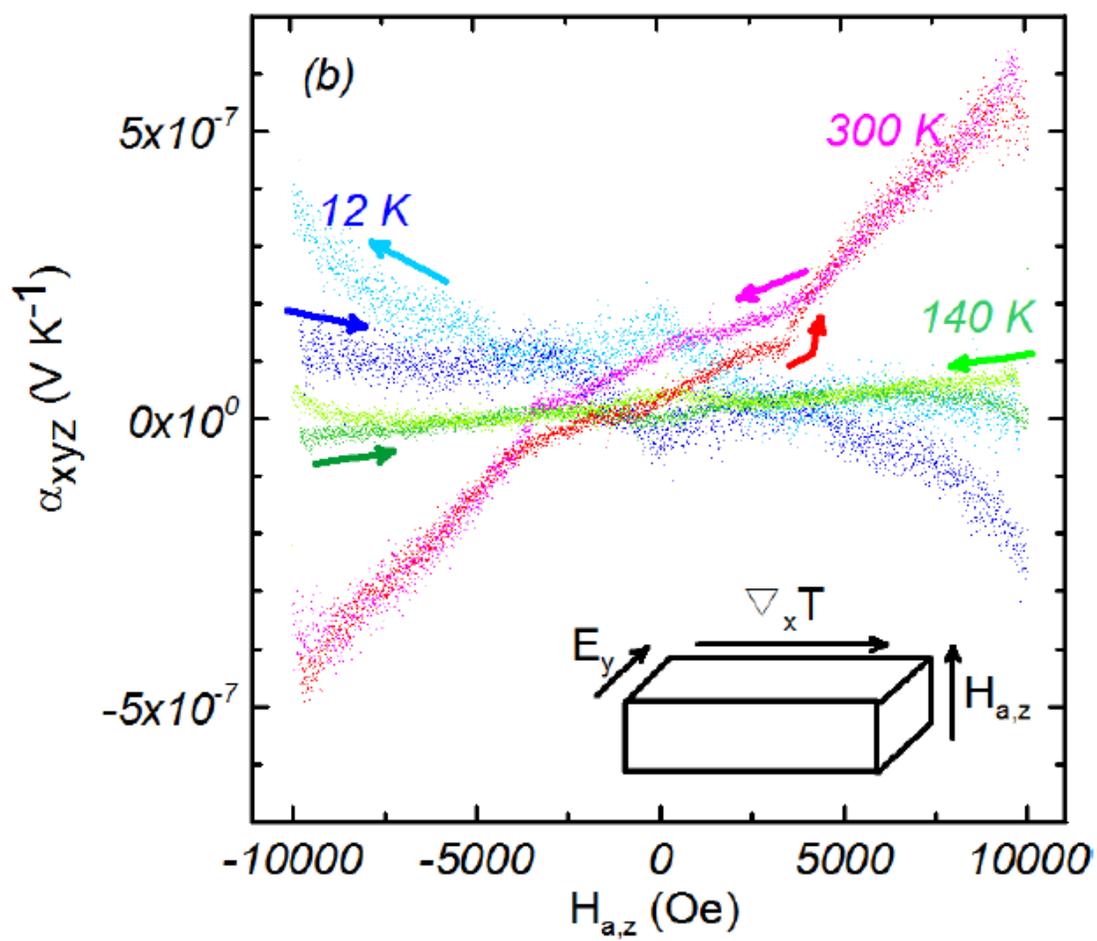

Fig. 3b

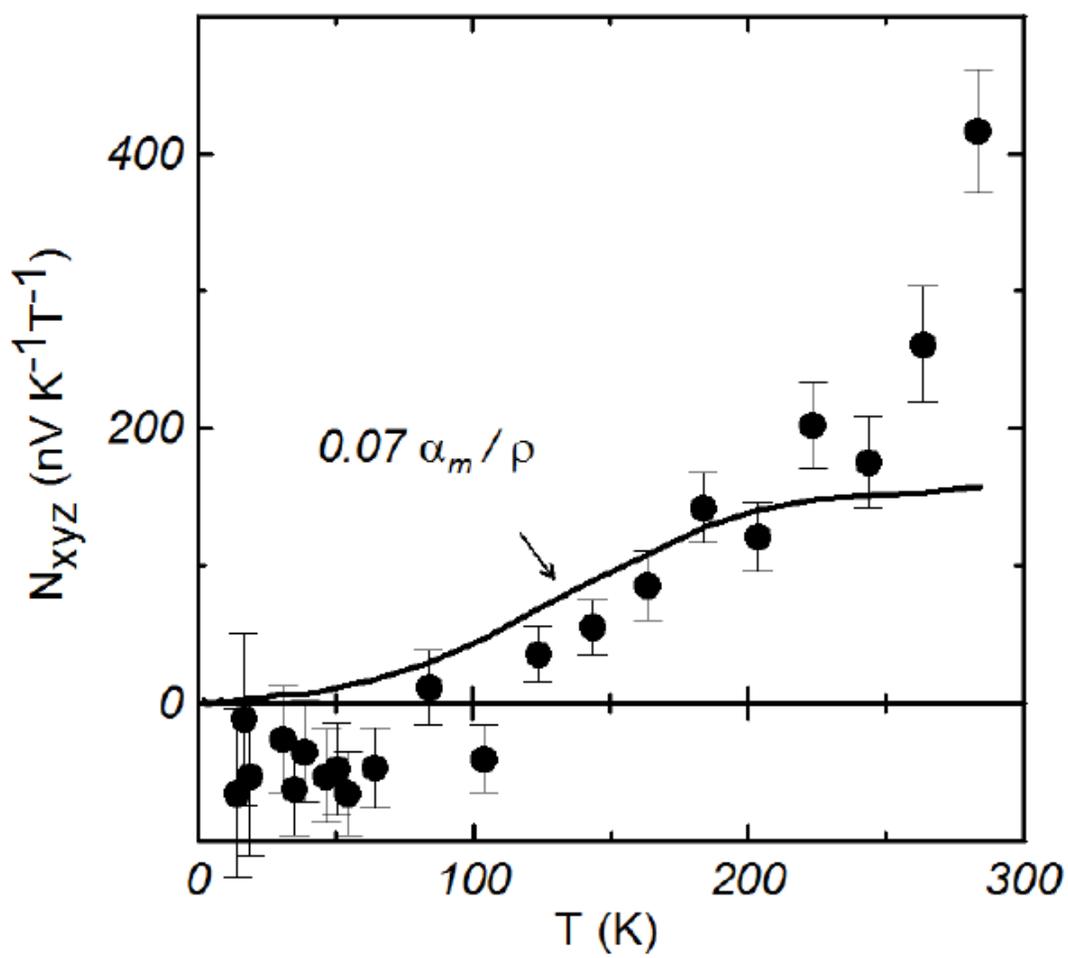

Fig. 4

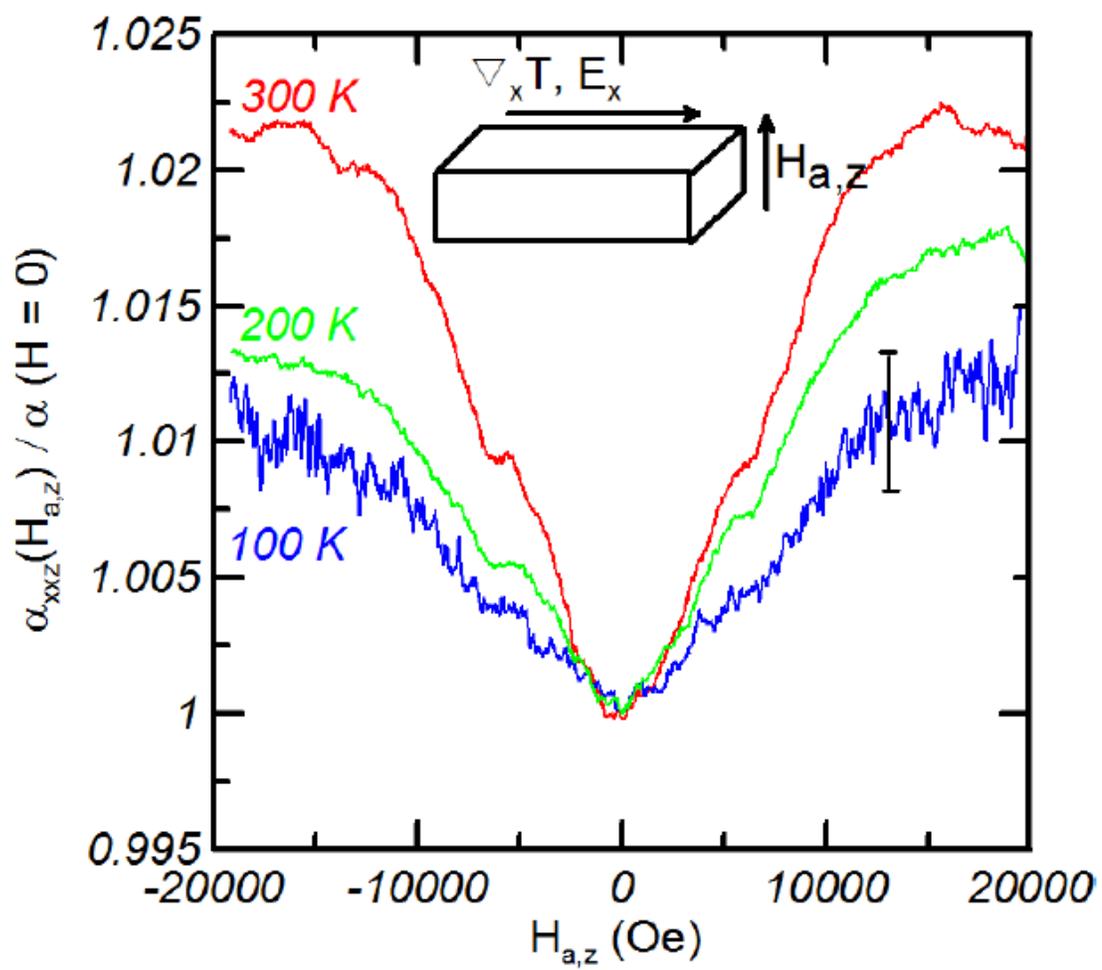

Fig. 5

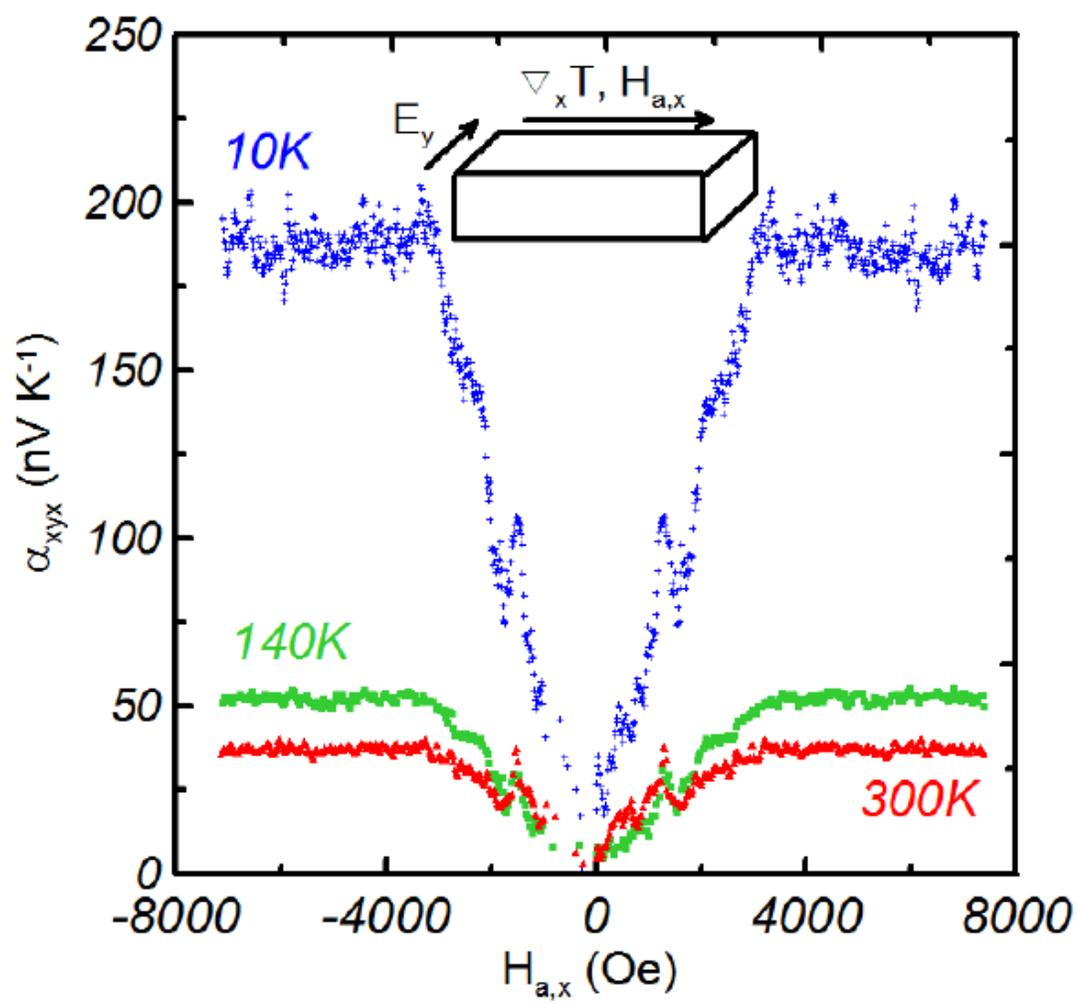

Fig. 6

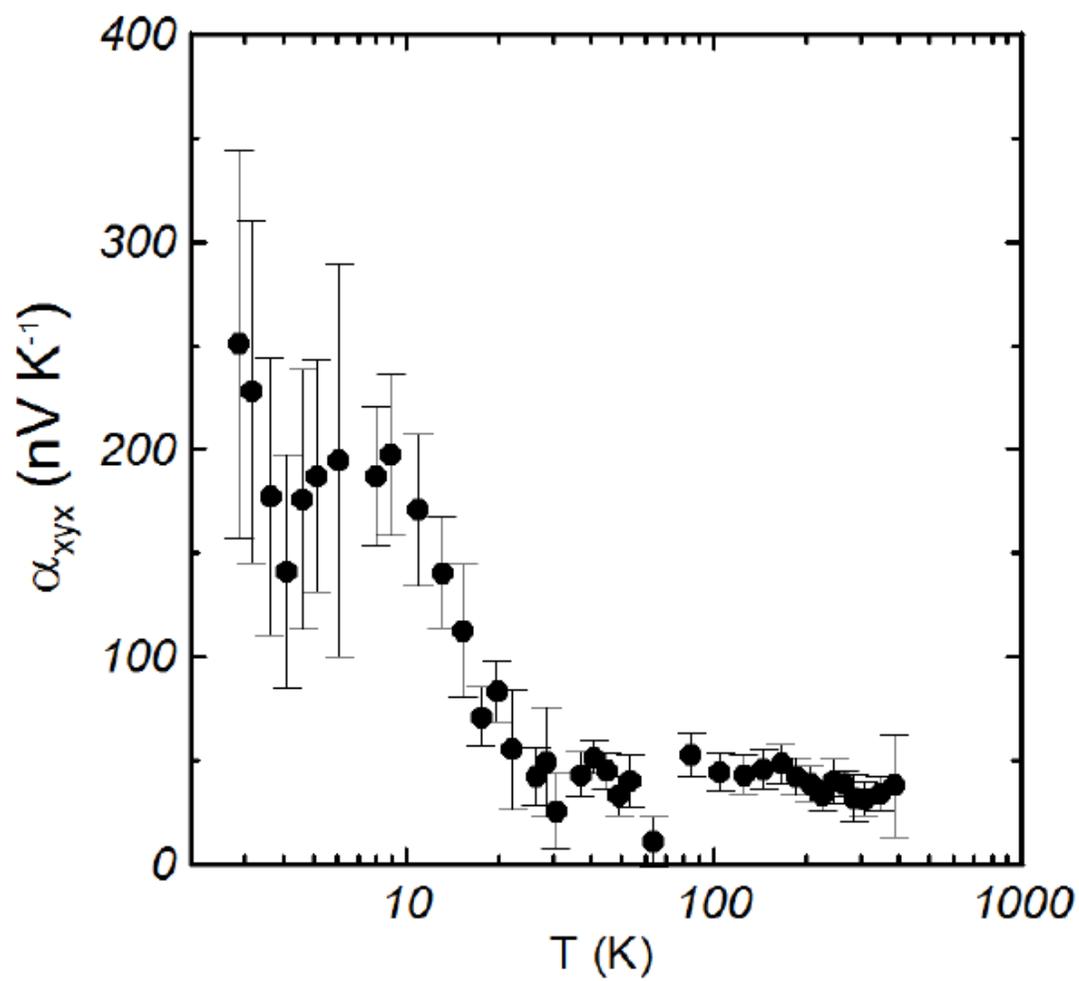

Fig. 7